\documentclass[11pt]{article}

\usepackage{latexsym}
\usepackage{algorithm}
\usepackage[xdvi,dvips]{graphics}
\usepackage{pstricks}
\usepackage{epsfig}
\usepackage{amssymb}
\usepackage{multicol}
\usepackage{array}
\usepackage{float}
\graphicspath{{figures/}{./}}

\newenvironment{proof}{{\bf Proof. } }{{\hfill $\Box$}\vspace{.5pc}}
\newtheorem{theorem}{Theorem}[section]

\newtheorem{definition}[theorem]{Definition}
\newtheorem{lemma}[theorem]{Lemma}

\newtheorem{remark}[theorem]{Remark}

\newtheorem{proposition}[theorem]{Proposition}

\floatstyle{ruled}

\newfloat{algo}{thp}{lop}[section]
\floatname{algo}{Algorithm}

\newcommand{\BEGLIST}{\begin{list}{}{\partopsep -2pt \parsep -2pt \listparindent 0pt}}% \labelwidth .5in}}
\newcommand{\ENDLIST}{\end{list}}

\newcommand{\mN}   {\mathcal{N}}

\newcommand{\mP}   {\mathcal{P}}

\newcommand{\mX}   {\mathcal{X}}
\newcommand{\bbZ}   {\mathbb{Z}}

\newcommand{\TRUE}{\mathtt{true}}
\newcommand{\FALSE}{\mathtt{false}}

\newcommand{\text}[1]{\mbox{#1}}

\begin{document}

%\title{  Self-Stabilizing $\Delta$-Barrier Synchronization  
%in Asynchronous Anonymous Networks}

\title{ Unison as a Self-Stabilizing Wave Stream Algorithm 
in Asynchronous Anonymous Networks}
%\small{Extended Abstract}

%\title{ Self-Stabilizing $\Delta$-Wave Stream  and  $\Delta$-Barrier Synchronization  
%in Asynchronous Anonymous Networks}

\author{
Christian Boulinier\\
LaRIA, CNRS FRE 2733\\
Universit\'{e} de Picardie Jules Verne, France\\
}

\date{}
\maketitle
\footnotesize
\begin{abstract}
How to pass from  local to  global scales in anonymous networks? In such networks, how to organize a self-stabilizing  propagation of information with feedback? From 
Angluin's results, the deterministic leader election is impossible in
 general anonymous networks. Thus, it is impossible to build a rooted
spanning tree. In this paper we show how to use Unison to
 design a self-stabilizing \emph{barrier synchronization} in an anonymous network. We show that the communication structure of this barrier synchronization designs a  self-stabilizing wave stream, or pipelined wave, in anonymous networks.  
%According to Tel result, 
%the knowledge of an upper bound of the network diameter is necessary.
%This approach raises concurrency. A process can start a new wave, whill the  previous wave is not terminated. 
We introduce two variants of waves:  Strong Wave and Wavelet. 
Strong waves can be used to solve the idempotent $r$-operator parametrized  problem, which implies well known problems like depth-first search tree construction -- this instance requires identities for the processors.  Wavelets deal with $\rho$-distance computation. We show how to use Unison to design a self-stabilizing  strong wave stream,  and wavelet stream respectively.
%We  define barrier Synchronization at distance $k$, and local mutual exclusion at distance $k$, and we show how to solve  the self-stabilizing version of this two problems with Unison.

\end{abstract}
\normalsize

\thispagestyle{empty}

\baselineskip 5.0mm \-\unitlength=1mm

\bigskip {\bf Correspondance}:

\begin{quotation}
Christian BOULINIER

Email: Christian.Boulinier@u-picardie.fr

LaRIA, Universit\'{e} de Picardie Jules Verne, Amiens, France

Tel. +33-322-809-577

\end{quotation}

\newpage 

\setcounter{page}{1}

\section{Introduction}

%\cite{ABDT98}  \cite{Gar03} \cite{HL01}

Several general message passing problems are useful to achieve many tasks in  distributed networks, like broadcasting information, global
synchronization, reset, termination detection, or calculation of a global function  whose the
input depends on several processes or the totality of  the  processes in the
network -- see~\cite{RH90,Tel94,Lyn96}. 
%Two general tasks are considered as important,  \emph{leader election}
%and  \emph{wave propagation}. Following~\cite{Ang80}, a deterministic  \emph{leader
%election} algorithm is impossible in a general anonymous network. 
In this paper we consider the \emph{wave propagation} problem in asynchronous
anonymous networks.  

\subsection{Related Works}

%An interesting question is to know if $\Omega(D|E|)$ is a lower bound on the message complexity of decentralized anonymous wave algorithm.  \cite{As96} proves this result 
%for the unidirectional ring, the hypercube and the clique. 

In asynchronous systems, there is no global signal. Synchronization is a crucial task. Informally, a synchronizer allows asynchronous systems to simulate
synchronous ones.  
%This problem admits some variants: 
%centralized daemon, a network with indentified processes, or anonymous networks. 
In asynchronous systems, one can at most ensure that no process starts to execute 
its phase $i+1$ before all processes have completed their phase $i$. This strongest synchronization task, named \emph{Barrier Synchronization}, was introduced by Misra in~\cite{M91} in a complete graph. The research about synchronization started with Awerbuch~\cite{A85}. 
Communications waves are often used to achieve synchronization. 
Designing efficient fault-tolerant wave algorithms is an important task.
Self-stabilization~\cite{D74,D00} is a general technique to design a system that tolerates arbitrary
transient faults, i.e. faults that may corrupt the state of processes or links.
\cite{KA98} proposes a self-stabilizing solution for complet graphs. \cite{HL01} designs a solution in uniform rings with an odd size.
A relaxed synchronization requirement is defined as follows: 
the clocks are in phase if the values of two neighboring processes  differ by no more than $1$, 
and the clock value of each process is incremented by $1$ infinitely often.  
The \emph{self-stabilizing asynchronous unison}~\cite{CFG92} deals with this criterium. % -- see also~\cite{Her00}. 

A distributed protocol is \emph{uniform} if every process with the same degree executes the same program. 
In particular, we do not assume a unique process identifier -- the network is anonymous -- or some consistent orientation of links in the 
network such that any dynamic election of a master clock can be feasible.
Numerous self-stabilizing wave algorithms use a rooted spanning tree or simply an only initiator, called the \emph{root}-- see for instance \cite{K79} \cite{ABDT98}.% \cite{BDPV99b}.   
In these cases,  protocols are not uniform, they are only at most semi-uniform. So, for a uniform distributed protocol
any processor may  initiate a wave, and most generally a global computation. Any processor may be an initiator. 
To face this inherent concurency, a solution is that every processor maintains the identity of the initiators -- see for instance \cite{CDPV02}. 
That is impossible in an anonymous network.

\cite{KA98} designs a self-stabilizing Barrier Synchonization algorithm in asynchronous anonymous complet networks. For the other topologies
the authors use the network with a root,  the program is not uniform, but only semi-uniform.
An interesting question is to give a solution to this problem in a general connected asynchronous anonymous network. 
%A related question araises naturally:  How to design a self-stabilizing wave algoritm in such  a network.
As far as we know, the \emph{phase} algorithm \cite{Tel91} is the only  decentralised uniform wave algorithm for a general anonymous
network. This algorithm requires that the processors know the diameter, or
most simply a common upper bound $D^{\prime }$ of the diameter. % During a
%phase each process sends $D^{\prime }$ messages to its neighbors. Its message complexity is $O(D'|E|)$, where $|E|$ is the number of edges of the network. 
This algorithm is not self-stabilizing.

\subsection{Contribution and paper outline}

The main task of this paper is to show how Unison can be viewed as a%
\emph{\ self-stabilizing wave stream} algorithm in asynchronous anonymous
networks scheduled by an \emph{unfair daemon}. The contribution is threefold:

Firstly, we introduce the $\rho$-distance barrier synchronization notion. It is a small extention of the barrier synchronization~\cite{M91} which ensures that no process starts to execute 
its phase $i+1$ before all processes at distance less than or equal to $\rho$ have completed their phase $i$. We show how to design a self-stabilizing \emph{barrier synchronization} at distance $\rho$ in an anonymous network. The self-stabilizing time complexity is in $O(n)$ rounds.  It has its space complexity  in $O(log(n)+log(K))$, where $n$ is the number of processes in the network and $K$ the size of the clock.
Secondly, we introduce two variants of Wave: Wavelet and Strong Waves. 
We show that a strong wave can be used to solve the idempotent $r$-operator parametrized  problem, and a wavelet deals with $\rho$-distance computation. 
Thirdly, we show that the communication structure of our $\rho$-distance barrier synchronization designs a  self-stabilizing wavelet stream, or pipelined wavelet, in any anonymous networks.  We show that  if $\rho \geq D$ the communications   design a self-stabilizing wave stream , and if $\rho $ is greater than or equal to the length of the longest simple path in the network, then the protocol designs a self-stabilizing strong-wave stream.

The remainder of the paper is organized as follows. 
In the next section (Section~\ref{sec:prel}), we describe the underlying model for distributed 
system. We  also state what it means for a protocol to be 
self-stabilizing, we introduce the notion of causal-$DAG$ and we present the  unison problem and its solutions.
In Section~\ref{sec:bar_syn} we define the  $\rho$-distance barrier synchronization notion and we introduce a protocol which designs a self-stabilizing \emph{barrier synchronization} at distance $\rho$ in any anonymous networks.  
In Section~\ref{sec:wave} we define two kinds of waves: 
 \emph{wavelet}  and \emph{strong waves}, and we show the  relationship between a strong wave and  the idempotent
$r$-operator parametrized computation problem .
In Section~\ref{sec:applications},  we show how Unison can be view as a wave stream, or a wavelet stream, or a strong wave stream.
In Section~\ref{sec:conclusion}, we give some concluding remarks. Because of the lack of place, somme proofs are put back in an annexe.

\section{Preliminaries}
\label{sec:prel}

In this section, firstly we define the model of distributed systems 
considered in this paper, and state what it means for a protocol to be
self-stabilizing.  Secondly, we present the notions of finite incrementing system
and reset on it. Next, we define what  a self-stabilizing distributed Unison is.\\

\subsection{The model}
\paragraph{Distributed System.}

A \emph{distributed system} is an undirected connected graph, $G=(V,E)$,
where $V$ is a set of nodes---$|V|=n,\ n \geq 2$---and $E$ is the set of edges. Nodes
represent \emph{processes}, and edges represent \emph{bidirectional
communication links}. A communication link $(p,q)$ exists iff $p$ and
$q$ are neighbors.
The set of neighbors of every process $p$ is denoted as $\mN_p$.
The \emph{degree} of $p$ is the number of neighbors of $p$, i.e., equal to $|\mN_p|$.
The distance between two processes $p$ and $q$, denoted by $d\left( p,q\right) $,
is the length of the shortest path between $p$ and $q$.  Let $k$ be a positive integer. Define $V(p,k)$ as the set of processes 
such that $d(p,q) \leq k$. $D$ is the diameter of the network.
%following subset of processes: be the If $k$ is an integer, we define 
%$D\left( p,k\right)=_{def}\left\{ q\in V,d\left( p,q\right) \leq k\right\} $. 
%The diameter $D$ of the network is equal to 
%$\max_{p\in V}\max_{q\in V}d(p,q)$.

The program of a process consists of a set %$\mR_p$ 
of registers (also referred to as variables) 
and a finite set %$\mA_p$ 
of guarded actions of the following form: 
$<label>::\ <guard>\ \longrightarrow <statement>$. 
Each process can only write to 
its own registers, and read its own registers and registers owned by the neighboring processes.
The guard of an action in the program of $p$ is a boolean
expression involving the registers of $p$ and its neighbors. The
statement of an action of $p$ updates one or more registers of $p$. 
An action can be executed only if its guard evaluates to true.   
The actions are atomically executed, meaning the evaluation of a guard and the execution of
the corresponding statement of an action, if executed, are done in one atomic step.
The \emph{state} of a process is defined by the values of its registers.
The \emph{configuration} of a system is the product of the states of all processes. 
%In the sequel, we refer to the state of a process and system as a
%(\emph{local}) \emph{state} and \emph{configuration}, respectively. 
Let a distributed protocol $\mP$ be a collection of binary transition
relations denoted by $\mapsto $, on $\mathcal{C}$, the set of all
possible configurations of the system.  $\mP$ describes an
oriented graph $S=(\mathcal{C}, \mapsto)$, called the \emph{transition graph} of $\mP$.
A sequence $e=\gamma_0, \gamma_1, \ldots, \gamma_i,\gamma_{i+1},\ldots$ is called an
\emph{execution} of $\mP$ iff $\forall i\geq 0, \gamma_{i}\mapsto \gamma _{i+1} \in S$. 
A process $p$ is said to be \emph{enabled} in a configuration
$\gamma_i \;( \gamma _i\in \mathcal{C})$ if there exists an action $A$ such that 
the guard of $A$ is true in $\gamma_i$.  The value of a register $r$ of a process $p$ in the state $\gamma_i$, is denoted by $p^i.r$. $i$ is the moment of the state $\gamma_i$.
When there is no ambiguity, we will omit $i$.
Similarly, an action $A$ is said to be enabled (in $\gamma$) at $p$
if the guard of $A$ is true at $p$ (in $\gamma$).
We assume that each transition from a configuration to another is driven by
a \emph{distributed scheduler} called \emph{daemon}. 
In this paper, we consider only an Asynchronous distributed Daemon. The \emph{Asynchronous Daemon} chooses any nonempty set of enabled 
processes to execute an action in each computation step (Unfair Daemon).

%If is not required to be \emph{fair}, i.e., even if a 
%process $p$ is continuously enabled, then $p$ may never be chosen by 
%the daemon unless $p$ is the only enabled process.

In order to compute the time complexity, we use the definition of
\emph{round}~\cite{DIM97a}.  This definition captures the execution rate of 
the slowest processor in any computation.
Given an execution $e$, %($e \in \mathcal{E}$)
 the \emph{first round} of $e$ 
(let us call it $e^{\prime}$)
is the minimal prefix of $e$ containing the execution of one action 
of the protocol or the neutralization of every enabled processor from the first configuration.  
Let $e^{\prime \prime}$ be the suffix of $e$, i.e., $e=e^{\prime}e^{\prime \prime}$.  
Then \emph{second round} of $e$ is the first round of $e^{\prime \prime}$, and so on.

%The distributed systems considered in this paper are assumed to be uniform.
%A distributed protocol is \emph{uniform} if every process with the same degree executes the same program.
%In particular, we do not assume a unique process identifier or some consistent orientation of links in the 
%network such that any dynamic election of a master clock can be feasible. 

\paragraph{Self-Stabilization.}
Let $\mX$ be a set. A \emph{predicate} $P$ is a function that has a Boolean 
value---$\TRUE$ or $\FALSE$---for each element $x\in \mX$.
A predicate $P$ is \emph{closed} for a transition graph $S$ iff 
every state of an execution $e$ that starts in a state satisfying $P$ also satisfies $P$.
A predicate $Q$ is an attractor of the predicate $P$, denoted by $P \vartriangleright Q$,
iff $Q$ is closed for $S$ and for every execution $e$ of $S$, beginning by a state satisfying $P$, 
there exists a configuration of $e$ for which $Q$ is true. 
A transition graph $S$ is \emph{self-stabilizing} for a predicate $P$ iff $P$ is an attractor 
of the predicate $\TRUE$, i.e., $\TRUE \vartriangleright P$.

\subsection{Causal DAGs}

\begin{definition}[Events and Causal DAGs] 
Let $\gamma _{t_{0}}\gamma _{t_{0}+1}....$ be a finite or
infinite execution. 
 $\forall p\in V,\left( p,t_{0}\right) $ is an event.
  Let $\gamma _{t}\rightarrow \gamma _{t+1}$ be a transition. If the
process $p$ executes a guarded action during this transition, we say that $p$
executes an action at time $t+1$, and we say that $\left( p,t+1\right) $is an
event or a $p$-event. 
%An \emph{internal event} is an event such that the guard  does not depend on the shared registers of any neighbor.
The \emph{causal DAG} associated is the smallest relation $\leadsto $ on the set of events
that satisfies:
\begin{enumerate}
\item  Let $\left( p,t\right) $ be an event with $t>t_{0}$.  Let $t^{\prime }$
be the greatest integer such that $t_{0}\leq t^{\prime }<t$ and $\left(
p,t^{\prime }\right) $ is an event, then $\left( p,t^{\prime }\right)
\leadsto \left( p,t\right) $

\item  
Let $(p,t)$ be an \emph{event} %which is not an \emph{internal event}
and let $t>t_0$. Let $q\in 
\mathcal{N}_{p}$ and let $t^{\prime }$ be the greatest integer such that $t_{0}\leq t^{\prime }<t$ and such that $\left( q,t^{\prime }\right) $ is an event,
then $\left( q,t^{\prime }\right) \leadsto \left( p,t\right) $.
\end{enumerate}

The \emph{causal order} $\preceq $ on the set of events is the reflexive and transitive closure of the causal
relation $\leadsto $.
The \emph{past cone} of an event $\left( p,t\right) $ is the  causal-$DAG$ induced
by every event $\left( q,t^{\prime }\right) $ such that $\left( q,t^{\prime
}\right) \preceq \left( p,t\right) $. 
A \emph{past cone} involves a process $q$ iff there is a $q$-event in the cone.
 The cover of an event $(p,t) $ is the set of processes $q$ covered by the past cone of $(p,t)$, this set is denoted by $Cover(p,t)$. 

\end{definition}

\begin{definition}[Cut]
A cut $C$ on a causal DAG is a map from $V$ to $\Bbb{N}$, which associates
each process $p$ with a time $t_{p}^{C}$ such that $(p,t_{p}^{C})$ is an event. We mix this map with its graph: 
$C=\left\{ \left( p,t_{p}^{C}\right) ,p\in V\right\} $.
The past of $C$ is the
events $(p,t)$ such that $t\leq t_{p}^{C}$. It is denoted by $\left]
\leftarrow ,C\right] $. %\newline
The future of $C$ is the events $(p,t)$ such that $t_{p}^{C}\leq t$. It is
denoted by $\left[ C,\rightarrow \right[ $.%\newline
A cut is \emph{coherent} if $\left( q,t^{\prime }\right) \preceq \left(
p,t\right) $ and $\left( p,t\right) \preceq \left( p,t_{p}^{C}\right) $
then $\left( q,t^{\prime }\right) \preceq \left( q,t_{q}^{C}\right) $.
%\newline
A cut $C_{1}$ is less than or equal to a cut $C_{2}$, denoted by $C_{1}\preceq C_{2}$,
if the past of $C_{1}$ is included in the past of $C_{2}$.%\newline
If $C_{1}$ and $C_{2}$ are coherent and $C_{1}\preceq C_{2}$ then $\left[ C_{1},C_{2}\right] $ is the 
\emph{induced} causal DAG defined by the events $\left( p,t\right) $ such
that $\left( p,t_{p}^{C_{1}}\right) \preceq \left( p,t\right) \preceq \left(
p,t_{p}^{C_{2}}\right) $.%\newline
Any segment $\left[ C_{1},C_{2}\right] $ is a  \emph{sequence of events}, each event of $C_1$ is called an \emph{initial event}.
\end{definition}

\subsection{Distributed Unison}
\label{sec:unison}

\emph{Unison}, or most precisely \emph{Self-Stabilizing Asynchronous Unison}, is a relaxed 
\emph{self-stabilizing Barrier Synchronization} in the following meaning: 
the clocks are in phase if the values of two neighboring processes  differ by no more than $1$, 
and the clock value of each process is incremented by $1$ infinitly often. 
Self-stabilizing Unison was introduced by \cite{CFG92}. 
% Let WU the set of states where two neighboring processes  differ by no more than 1. It is easy to show that any guarded incrementing action,  for which WU is closed, is equivalent to the guarded action~(\ref{eq:guarde}), next page. 
There is a possibility of deadlock if the size of the clock is too short-- see\cite{BPV04b}.  %Moreover, if we want to stabilize Unison with a reasonable time complexity, we must use an incrementing system. 
A little algebraic framework, and some vocabulary  are necessary. The vocabulary will be used in the definition of the  algorithm~\ref{algo:SSWS}. 
~%Following \cite{BPV04b}, we briefly introduce the important notion of \emph{intrinsic delay}, and the notion of \emph{cyclomatic characteristic}. Then, we define Unison problem and we give  the state of art about it.

\paragraph{Algebraic framework} Let $\bbZ$ be the set of integers and $K$ be a strictly positive integer.
Two integers $a$ and $b$ are said to be \emph{congruent modulo} $K$, denoted by
$a\equiv b [K]$ if and only if $\exists \lambda \in \bbZ,\
b = a + \lambda K$. We denote $\bar{a}$ the unique element in $[0,K-1]$
such that $a \equiv \bar{a} [K]$.  $\min ( \overline{a-b}, \overline{b-a})$ is a \emph{distance} 
on the torus $[0,K-1]$ denoted by $d_K (a,b)$ . 
Two integers $a$ and $b$ are said to be \emph{locally comparable}  if and only if
 $d_K (a,b) \leq 1$.  We then define
the \emph{local order relationship} $\leq_l$ as follows: 
$
a\leq_l b \stackrel{\mathrm{def}}{\Leftrightarrow} 0 \leq \overline{b-a} \leq 1
$.
If $a$ and $b$ are two locally comparable integers,  we define  
$b \ominus a $ as follows:
$
b \ominus a =_{def}\mbox{  if } a \leq _l b \mbox{ then } \overline{b-a} \mbox{ else }  - \overline{a-b}    
$.
%Note that $b \ominus a \equiv b-a [K]$.  So, 
If $a_0, a_1, a_2,\ldots a_{p-1}, a_p$ is a sequence of integers such that 
$\forall i \in \{0,\ldots,p-1\}$, $a_i$ is locally comparable to $a_{i+1}$,
then $S=\sum \limits_{i=0}^{p-1}\left( a_{i+1} \ominus a_i\right) $ is the \emph{local variation} of this sequence. 
%Clearly, $S \equiv a_p - a_0 [K]$.% and 
%$S \equiv 0 [K] \Leftrightarrow a_p - a_0 \equiv 0[K].$ 
 
\paragraph{Incrementing system} We define $\mX = \{-\alpha,\ldots,0,\ldots,K-1\}$, where $\alpha$ is a positive integer. 
Let $\varphi$ be the function from $\mX$ to $\mX$ defined by:
%$$ 
%\varphi: x \rightarrow 
%\left\{ 
%  \begin{array}{ll}
%     \overline{x+1} & \mbox{if } x \geq 0 \\
%     x+1            & \mbox{otherwise}
%  \end{array}
%\right. 
%$$
$
\varphi (x)=_{def}\mbox{  if } x \geq 0 \mbox{ then } \overline{x+1} \mbox{ else }  x+1    
$.
The pair $(\mX,\varphi)$ is called a \emph{finite incrementing system}. %---refer to Figure %~\ref{fig:cerise}.
%
%\begin{figure}[!htbp]
%\begin{center}
%    \epsfig{file=cerise.eps, width=0.35\linewidth}\\
%\end{center}
%\caption{The finite incrementing system $(\mX,\varphi)$.}
%\label{fig:cerise}
%\end{figure}
$K$ is called the \emph{period} of $(\mX, \varphi)$.
%Let $\alpha = -\kappa$ be the \emph{initial element} of $(\mX,\varphi)$.
Let $tail_{\varphi}=\{-\alpha,\ldots,0\}$ and $stab_{\varphi}=\{0,\ldots,K-1\}$ be the sets of 
``extra'' values and ``expected'' values, respectively.
The set $tail_\varphi ^*$ is equal to $tail_\varphi \setminus \{0\}$.
%We denote by $\leq _{tail}$ the natural total order on $tail_\varphi$, and $\leq$ the natural
%order on $\mX$.
%A \emph{reset} on $\mX$ consists in enforcing any value of $\mX $ to $-\alpha$.  
%
%In this section, we briefly state some assertions around the asynchronous unison problem and its solutions.  
%
%
%\paragraph{ $WU$ predicat, and Path Delay }
%and Precedence Relation
%
We assume that each process $p$ maintains a clock register $r_p$ with an incrementing system $(\mX,\varphi)$. 
Let $\gamma $ the system configuration, we define the predicate $WU$:
 
%$WU(\gamma )\stackrel{\mathrm{def}}{\equiv} \forall p \in V,\forall q\in \mN_p:(r_p \in stab_{\varphi })\wedge (  | r_p - r_q | \leq 1)$ in $\gamma$.  
\small
$$
WU(\gamma )\equiv_{def} \forall p \in V,\forall q\in \mN_p:(r_p \in stab_{\varphi })\wedge (  d( r_p , r_q ) \leq 1)  \text{ in } \gamma .  
$$
\normalsize
%In the remainder, we will abuse notation, referring to the corresponding set of 
%configurations simply by $WU$. Now, let us recall the notion of path delay and some of its properties established in \cite{BPV04b}.  

\paragraph{Intrinsic Path Delay \cite{BPV04b}}
Let $\gamma$ a  configurations in $WU$, the clock values of neighboring processes are locally comparable.  We define the four notions: 
%\begin{enumerate}
%\item 

%\small
\textbf{Delay}  \label{def:delay} The delay along a path $\mu = p_0p_1\ldots p_k$, denoted by
$\Delta _\mu$, is the local variation of the sequence 
$r_{p_0}, r_{p_1},\ldots, r_{p_k}$, 
i.e, $\Delta _\mu = \sum \limits_{i=0}^{k-1}\left( r_{p_{i+1}} \ominus _l r_{p_i}\right)$ if 
$k > 0$, $0$ otherwise ($k=0$).
%\item 
 
\textbf{Intrinsic Delay} \label{def:intrinsic}
The delay between two processes $p$ and $q$ is \emph{intrinsic} if it is independent on the choice
of the path from $p$ to $q$.  The delay is \emph{intrinsic} iff it is \emph{intrinsic} for every 
$p$ and $q$ in $V$. In this case, and at time $t$, the intrinsic delay between $p$ and $q$ is denoted by $\Delta_{(p,q)}$.
%\item 

\textbf{WU$_0$} The predicate $ WU_0$ is true for a system configuration $\gamma$ iff $\gamma$ satisfies $WU$ 
and the delay is intrinsic in $\gamma$. 

\textbf{Precedence relationship}
\label{rem:delay}
When Delay is intrinsic, it defines a total preordering  on the processes in $ V$, named \emph{precedence relationship}. This relationship depends on the state $\gamma \in WU_0$. The absolute value of the delay between two processes  $p$ and $q$, 
is equal to or less than 
the distance $d(p,q)$ in the network. This remark is important for the following.
 
\paragraph{Cyclomatic Characteristic $C_G$ \cite{BPV04b}}
\label{def:CG}
If $G$ is an acyclic graph, then its cyclomatic characteristic $C_G$ is equal $2$. 
Otherwise $G$ contains cycles:  Let $\Lambda$ be a cycle basis,  the length of the longest 
cycle in $\Lambda $ is denoted $\lambda (\Lambda)$.
The cyclomatic characteristic of $G$,  is equal to the lowest 
$\lambda ( \Lambda ) $ among cycle bases. It follows from the definition of  $C_G$ that $C_G \leq 2D$.

%Let us define Predicate $WU_0$ which is true for a system configuration $\gamma$ iff $\gamma$ satisfy $WU$ 
%and the delay is intrinsic in $\gamma$.  
%Clearly, the residuals are congruent to $0$ modulo $K$. 
%The delay being intrinsic iff it is equal to $0$ on every cycle, by linearity, 
%a path delay is intrinsic iff it is equal to $0$ on a cycle basis \cite{B89}.  
%We borrow the following definition from \cite{BPV04b}:

\paragraph{Unison Definition}
We assume that each process $p$ maintains a register $p.r \in \chi$. 
The self-stabilizing \emph{asynchronous} (\emph{distributed}) \emph{unison} problem, or most shortly the \emph{unison } problem, is 
to design a uniform protocol so  that the following properties are true in every execution~\cite{BPV05}:  

\textbf{Safety }: $WU$ is closed.
\textbf{Synchronization}:  
In $WU$, a process can increment its clock $r_p$ only if the value of $r_p$ is lower than or equal to
the clock value of all its neighbors. 
%, i.e., $\forall q \in \mN_p,\ r_p \leq_l r_q$; 
\textbf{No Lockout (Liveness)}: 
In $WU$, every process $p$ increments its clock $r_p$ infinitely often. 
\textbf{Self-Stabilization }:  
$\Gamma \triangleright WU$.

The following guarded action solves the \emph{synchronization property} and the \emph{safety}:
\small
\begin{equation}
\label{eq:guarde}
\forall q \in \mN_p:\ (r_q=r_p)\vee (r_q =\varphi (r_p))\longrightarrow r_p := \varphi(r_p);
\end{equation}
\normalsize
%Remark that If $K \geq 4$ and the safety is satisfied, then the synchronization is satisfied. 
 The predicate
$WU_0$ is closed for any execution of this guarded action.  Moreover,
for any execution starting from a configuration in $WU_0$, the \emph{no lockout property} is
guaranteed. Generally this property is not guaranteed in $WU$.
%If a uniform distributed protocol $\mP$ solves the asynchronous distributed unison problem, then any action in $WU$ is functionaly equivalent to 
%the called Normal  Garded Instruction:
%$$
% \forall q \in \mN_p:\ (r_q=r_p)\vee (r_q =\varphi (r_p)) \longrightarrow r_p := \varphi(r_p);
%$$
%\end{lemma}
%Conversely, \cite{BPV04b}   shows that if $K>C_G$ then the above guarded action solves the asynchronous unison problem in $WU$. 
%If $K\leq C_G$, there is a possibility of deadlocks~\cite{BPV04b}.
A few general schemes to self-stabilizing the non-stabilizing protocols have been proposed. 
The first self-stabilizing asynchronous unison was introduced in~\cite{CFG92}.  The deterministic protocol 
 proposed needs $K \geq n^2$. The stabilization time complexity is in $O(nD)$.
The second solution is proposed in \cite{BPV04b}.  The authors show that  if  $K$ is
greater than $C_G$  then $WU=WU_0$ and the \emph{no lockout property} is
guaranteed in $WU$. 
(see Definition~\ref{def:CG}).
The protocol is self-stabilizing if   $\alpha \geq T_G-2$, where $T_G$ is the length of 
the longest chordless cycle ($2$ in tree networks).
One can notice that $C_G$ and $T_G$ are bounded by $n$.  So, even if $C_G$ and $T_G$ are unknown, 
we can choose $K \geq n+1$ and $\alpha=n$. Its self-stabilizing time complexity is in $O(n)$.
In \cite{BPV06}, the authors present the Protocol $WU\_Min$, which is 
self-stabilizing to asynchronous unison in at most $D$ rounds in trees.

\section{Barrier Synchronization}

\label{sec:bar_syn}
\subsection{Barrier synchronization at distance $\rho $}

\emph{Barrier Synchronization problem} has been specified in~\cite{KA98}. Let $\rho $ be an integer greater than $0$. The relaxation of this problem at distance $\rho $ is the following.
Let $K$ be an integer greater than $1$. We assume that each process $p$ 
 maintains a $K$-\emph{order} \emph{\ clock register} $p.R\in \left\{ 0,1,...,K-1\right\} $. 
Each process executes a cyclique sequence of $K$ terminating phases (the critical section $<<cs>>$). The following two properties are required for each phase:

\textbf{Global} \textbf{Unison (Safety) :} for each phase $x\in \left\{
0,...,K-1\right\} ,$\textbf{\ }no process $p$ can proceed to phase $\overline{x+1}$ until all nodes $q$, such that $d(p,q)\leq \rho $, has executed its phase $x$.

\textbf{No lockout (liveness)}: every process increments its clock infinitly
often.

For $\rho =1$, this specification is the specification of the standard stabilized unison. For $\rho \geq D$, this specification is the specification of the global Barrier Synchronization.

\subsection{The general self-stabilizing Scheme}

The idea is to stabilize an underlayer unison   
 in order to  synchronize a $\delta K-$clock,  with $\delta $ large enough
to guarantee that the absolute value of the delay between every two processes at distance less than or equal to $\rho$ is never larger than 
$\delta $. It is sufficient that $\delta \geq \rho$ holds. 
We take $\chi
=\left\{ -\alpha ,..,0,..,\delta K-1\right\}$ and $\alpha \geq T_G-2$ . 
We use the unison of \cite{BPV04b}  which  stabilizes in $O(n)$. The protocol is describe in Algorithm~\ref{algo:SSWS}.
To ensure self-stabilization in $WU_0$, we require $\delta K >C_G$. 
If we want to program a Barrier Synchronization, we must take $\delta \geq D$, thus from $C_{G}\leq 2D$, if $K\geq 3$ then
the inequality $K\delta >C_{G}$ holds. In the remainder we suppose that the inequality $K\delta >C_{G}$  holds.

\begin{algorithm}
\begin{footnotesize}
\noindent
{\bf Constant and variable}:\\
\hspace*{0.3cm}
$\mN_p$: the set of neighbors of process $p$; % \\
%\noindent
%\hspace*{0.3cm}
$p.r \in \chi $;\\
\noindent
{\bf Boolean Functions}:\\
\noindent
\hspace*{0.3cm}
\begin{tabular}{@{}lcl}
$ConvergenceStep_p$&$\equiv$ &$p.r \in tail_{\varphi }^{*}\wedge (\forall q\in \mN_p:(q.r \in tail_{\varphi })\wedge (p.r\leq _{tail_{\varphi }}q.r))$;\\
$LocallyCorrect_p$ &$\equiv$ & $p.r \in stab_{\varphi}\wedge (\forall q\in \mN_p,q.r \in stab_{\varphi}\wedge ( \left( p.r=q.r\right) \vee \left( p.r=\varphi \left(
q.r\right) \right) \vee \left( \varphi \left( p.r\right) =q.r\right) ))$;\\
$NormalStep_p$ &$\equiv$ & $p.r \in stab_{\varphi}\wedge (\forall q \in \mN_p:\ (p.r=q.r)\vee (q.r =\varphi (p.r)))$;\\
$ResetInit_p $ &$\equiv$ &
      $\neg LocallyCorrect_p\wedge (p.r \not\in init_\varphi)$;\\
\end{tabular}\\
{\bf Actions}:\\
\noindent
\hspace*{0.3cm}
\begin{tabular}{@{}rlcl}
 %R1
   $NA:$ &  
   $NormalStep_p$ & $\longrightarrow$ & 
 if $p.r\equiv \rho -1 [\rho ]$ then $ <<\mbox{ CS 2}>>$ else $<<\mbox{ CS 1}>>$ ; $p.r := \varphi(p.r)$;\\

%R2
   $CA:$ & %\ 
   $ConvergenceStep_p$ & $\longrightarrow$ &
   $p.r := \varphi(p.r)$;\\
%R3
   $RA:$ & %\ 
   $ResetInit_p$ & $\longrightarrow$ &
   $p.r := \alpha$ (reset);\\

\end{tabular}
\end{footnotesize}
\caption{($SS-WS$)Self-Stabilizing $\rho$-Barrier Synchronization algorithm  for the process $p$}
\label{algo:SSWS}
\end{algorithm}

\subsection{Analysis}

\paragraph{Lifting construction} In order to analyse the protocol~\ref{algo:SSWS} we introduce for each process $p$, a global device, the  register $\widetilde{p.r}$. 
Of course the value of this virtual register is  inaccessible to the process $p$. Informally $\widetilde{p.r}$ is a way to unwind of the register $p.r$.  Let $\gamma _{t_{0}}\gamma _{t_{0}+1}....$ be an  infinite execution 
starting in $WU_0$. Let $p_{0}$ be a maximal process, according to the \emph{precedence relation} -- see Remark~\ref{rem:delay} 
 -- for the state $\gamma _{0}$. Let $\bot _{0}=p_{0}.r$ at time $0$.
For each process $p\in V$, we unwind the register $p.r$ in the following manner. We
associate a virtual register $\widetilde{p.r}$. For the state $\gamma _{0}$,
we initiate this virtual register by  the instruction $\widetilde{p.r}:=\bot _{0}+\Delta
_{(p_{0},p)}^{0}$. During the execution, for each  transition $\gamma
_{t}\rightarrow \gamma _{t+1}$ the intruction $\widetilde{p.r}:=\widetilde{p.r}+1$ holds 
if and only if $p.r:=\overline{p.r+1}$ holds during the same transition.
For $k\geq \bot _{0}$ we define the cut $C_{k}=\left\{ \left(
p,t_{p,k}\right) ,p\in V\right\} $ where $t_{p,k}$ is the smallest time such
that $\widetilde{p.r}:=k$.
The first question is to prove that this cuts are coherent. We first introduce the easy lemma:

\begin{lemma}
\label{lem:coherent}
If $\left( p,t\right) \leadsto $ $\left( q,t^{\prime }\right) $ then: 
$
\widetilde{q^{t^{\prime }}.r}\in \left\{ \widetilde{p^{t}.r},\widetilde{p^{t}.r}+1\right\} 
$.
Inductively, if $\left( q_{0},t_{0}\right) \leadsto $
$\left( q_{1},t_{1}\right) \leadsto $ $\left( q_{2},t_{2}\right) ...\leadsto 
$ $\left( q_{i},t_{i}\right) $ then: 
$
\widetilde{q_{i}^{t_{i}}.r}\in \left\{ 
\widetilde{q_{0}^{t_{0}}.r},...,\widetilde{q_{0}^{t_{0}}.r}+i\right\}
$
\end{lemma}

From the Lemma~\ref{lem:coherent},  
if $\left( q,t\right) \preceq \left( p,t_{p,k}\right) $ then $\left( q,t\right)
\preceq \left( q,t_{q,k}\right) $. It follows the proposition:

\begin{proposition}
For every $k\geq \bot _{0}$ the cut $C_{k}$ is coherent.
\end{proposition}

\paragraph{ Virtual register $p.R$ and  virtual clock } For each process $p$ we associate the register $p.R$, which is virtual. Its value is evaluated by the procedure:  
if $p.r \in stab_{\varphi}$ then $p.R:= {p.r}/ \delta$ else $p.R:=-1$, 
where the symbol $/$ is the \emph{integer division} operator.  The virtual register $p.R$ defines a  clock  on $\left\{ -1,0,...,K-1\right\} $. The  algorithm~\ref{algo:SSWS} solves self-stabilizing Asynchronous Unison, so every process $p$
increments its clock $p.r$ infinitly often. We deduce that $p.R$ increments infinitly often, thus:

\begin{lemma}
(\textbf{Liveness}) For every process $p$, the virtual register $p.R$  is incremented infinitly often. Consequently $ <<\mbox{ CS 2}>>$ is
executed infinitly often. 
\end{lemma}

\begin{theorem}
If $\delta \geq \rho $, once the protocol is stabilized, it solves the
Barrier Synchronization at distance $\rho $ for the virtual clock defined by the register  $p.R$ .
\end{theorem}

\begin{proof}
We consider the phase $U=$ $\left[ C_{U\delta },C_{U\delta +\delta
-1}\right] $, for any event $(p,t)$ in this sequence, the register $p.R$ is equal to $%
\overline{U}\left[ K\right] $. Let $p$ and $q$ be two processes, such that $%
d(p,q)\leq \rho $. Let $(p,t_{p})$ and $\left( q,t_{q}\right) $ be in $%
C_{U\delta +\delta }$.\ Suppose that $t_{p}\leq t_{q}$,  at time $t_{p}$
the register $\widetilde{q.r}\in \left\{ U\delta +\delta -i,i\in \left\{
0,...,\rho -1\right\} \right\} $, thus at time $t_{p}$, the critical section $ <<\mbox{ CS 2}>>$ of the phase $U$ is
terminated for the process $q$.

\end{proof}

Our protocol synchronizes
processors at distance $\rho$ in any anonymous general network. 
On the general graph, this synchronizer does not need any identity and does not build any real or virtual
 spanning tree. 
%Usually, using a spanning tree, a communication wave consists of a broadcast
%from a root node to all other nodes, followed by a convergecast phase from
%the leaf nodes to the root node. 
Here, the broadcast runs in the beginning
of a phase from any decentralised node $p$. For each
node $q\in V(p,\rho)$, at the end  of the phase for $q$,  the node knows that
information is gone to all the others nodes in $ V(q,\rho)$,  the feedback is implicit. 
The time complexity of a phase $\left[ C_{U\delta },C_{U\delta +\delta
-1}\right] $ is $\delta $ rounds in worst case. The message
complexity is $2 \delta \left| E\right| $, which is the price to
pay for  uniformity. But is this message complexity  usable? We will give a positive answer.

\section{ Wavelet, Wave and Strong Wave}
\label{sec:wave}

During each phase of  Algorithm~\ref{algo:SSWS}, the structure of communications  is  a kind of wave depending of the value of $\delta$. These communication structures are formally defined in this  section.  In the section~\ref{sec:applications}, following the Theorem~\ref{th:unison_behavior},  we will be able to use these communications to compute some
important functions on the network, for instance an infimum if $\delta \geq D$, or most generally  the  idempotent $r$-operator parametrized calculation problem when $\delta \geq n$, 
and so to solve many silent tasks~\cite{Duc98}.

%In the first subsection, we define various general notions like \emph{%
%walk on a graph}, \emph{events, causal DAG and Causal ordering}. We define 
%\emph{Wave} notion in our model, we introduce the notion of \emph{Bidirected
%Link Flood.}
%In the second subsection we briefly introduce the two notions: \emph{infimum }and \emph{r-operator}.
%In the third subsection, following~\cite{Tel94}, we show the relationship between \emph{wave}
% and \emph{infimum calculation}. In the same spirit, we show the
%relationship between \emph{Bidirected Link Flood} and  the \emph{Idempotent
%r-operator Parametrized Calculation Problem }.
%In the last subsection, we show how to interpret Unison behavior as a \emph{Wave Stream }or as a \emph{Bidirected Link Flood Stream}.

\subsection{Walk and  Wave}

\begin{definition}
\textbf{Walk.} A \emph{Walk} is a finite non empty  word $m=q_{0}q_{1}.....q_{r}$ on the alphabet $V$, such that
for all $i\in \left\{ 0,r-1\right\}$, $q_{i}=q_{i+1}$ or $q_{i+1}\in \mathcal{N}_{q_{i}}$. 
 A walk is circular if $r>1$ and  $q_{0}=q_{r}$.  
The walk $m$ is beginning in $q_0$ denoted $head(m)$, and is ending in $q_r$. Its length is $r$.
%\end{definition}
%\begin{definition}[Reducing]

Let $m$ be a \emph{walk}, 
if there exists two words $m_{1}$ and $m_{2}$ , and a circular walk $u$ such that $m=m_{1}um_{2}$, ($u$ is a
factor of $m$), then $m_{1} head(u) m_{2}$ is a walk and we write:
$
m\rightarrow m_{1} head(u)m_{2}
$
%Where $\alpha $ is the \emph{beginning process} of $u$.
The transitive closure of  the relationship $\rightarrow $ defines a strict
partial ordering $\stackrel{*}{\rightarrow }$ in  the set of \emph{walks}. A \emph{%
simple walk} is a  \emph{minimal walk }according to the $\stackrel{*}{\rightarrow 
}$ partial ordering. Most simply, a \emph{simple walk} is a  \emph{walk} without any \emph{%
repetition}.
An \emph{elementary walk} is a \emph{walk} such that if for $i<j$ , $q_{i}=q_{j}
$ then for all $k\in \left\{ i,...,j\right\} $ , $q_{k}=q_{i}$.  A \emph{reducing} of a \emph{walk} $m$ is a \emph{simple walk} 
$m^{\prime }$ such that $m\stackrel{*}{\rightarrow }m^{\prime }$.

%Generally  a reducted walk is not unique, but if $m$ is \emph{elementary} then its reducing is unique. 
\textbf{Walk cover of an event in a sequence.} 
Let  $S=\left[ C_{1},C_{2}\right] $ be a \emph{sequence of events}. 
If in $S$, $\left( q,t^{\prime }\right) \preceq \left( p,t\right) $ then there
exists a \emph{causality chain }  from $\left( q,t^{\prime }\right) $ to $(p,t)$:  $\left( q,t^{\prime }\right) =\left(
q_{0},t_{0}\right) \leadsto $ $\left( q_{1},t_{1}\right) \leadsto $ $\left(
q_{2},t_{2}\right) ...\leadsto $ $\left( q_{r},t_{r}\right) =\left(
p,t\right) $ , its   associated \emph{walk} is the \emph{walk} $q_{0}q_{1}$...$q_{r}$. The \emph{walk cover} of an event $(p,t) \in S $ is the set of \emph{walks} associated to  the causality chains of $S$ ending to $(p,t)$.    This set is denoted by $WalkCover(p,t)$.  Of course, this set contains the \emph{walk} of length $0$ denoted by $p$. 
\end{definition}

\begin{lemma}
\label{lem:elt_walk_red}
If $m\in \emph{WalkCover(p,t)}$ then there exists an elementary walk $m^{\prime }$
 in $\emph{WalkCover(p,t)}$ such that $m\stackrel{*}{\rightarrow }m^{\prime }$ 
\end{lemma}

\begin{proof}
\small
Let $m=q_{0}q_{1}.....q_{r}$ the associated  walk of the causality chain $\left( q_{0},t_{0}\right) \leadsto $
 $\left( q_{1},t_{1}\right) \leadsto $ $\left( q_{2},t_{2}\right) ...\leadsto $ 
$\left( q_{r},t_{r}\right) $.
Suppose that  $m=m_{1}um_{2}$ where $u$ is a circular walk $q_{i}q_{{}}.....q_{j}$.
From the definition of $\leadsto $ relationship, there exists a chain: $%
\left( q_{i},t_{i}\right) \leadsto $ $\left( q_{i},t_{i_{1}}\right) \leadsto 
$ $\left( q_{i},t_{i_{2}}\right) ...\leadsto $ $\left( q_{i},t_{j}\right) $.
Let $l$ the length of this chain. If $v=\prod\limits_{k=1}^{l}q_{i}$ then $\bar{m}=m_{1}vm_{2}$   is an element of $\emph{WalkCover(p,t)}$. 
Such a \emph{rewriting operation} is possible only a finite number of times, at the end, the
word is elementary.
\normalsize
\end{proof}

\begin{definition}[Wavelet, Wave, and Strong Wave]
\label{def:wave}
Following~\cite{Tel94}, we assume that there are  special events called \emph{decide events}, the nature of these events depends of the algorithm. Let  $k$ an integer.
%\newline
A \emph{$k$-wavelet} is a \emph{sequence of events}  $\left[ C_{1},C_{2}\right] $
that satisfies the following two requirements:

%\begin{enumerate}
%\item  
The \emph{causal DAG }induced by $\left[ C_{1},C_{2}\right] $ contains at least one decide event. %notin $C_{1}$

%\item 
 For each \emph{decide event} $\left( p,t\right) $ %not in $C_{1}$
, the \emph{past } of $\left( p,t\right) $ in $\left[ C_{1},C_{2}\right] $ covers $
V(p,k).$\newline
%\end{enumerate}
We simply call it   a \emph{wave} when  $k \geq D$, where $D$ is the diameter of the network.
\newline
 A \emph{strong wave}  is a \emph{wave} $\left[ C_{1},C_{2}\right] $
that satisfies the following added requirement:

For each \emph{decide event}$\left( p,t\right) $ in $\left[ C_{1},C_{2}\right] $,  %wich is not in $C_{1}$
 and for each simple walk $m_0=q_{0}q_{1}$...$q_{n-1}p$ ending in $p$, there exists a 
\emph{causality chain }  $\left( q_{0},t_{0}\right) \leadsto $ $\left(
q'_{1},t_{1}\right) ...\leadsto $ $\left( q'_{r-1},t_{r-1}\right) \leadsto
\left( p,t\right) $ in $\left[ C_{1},C_{2}\right] $, such  that its associated walk $m$ is elementary, and  $m\stackrel{*}{\rightarrow }m_0$.

%\end{enumerate}

\end{definition}

%\begin{lemma}
%Let $\left[ C_{1},C_{2}\right] $ be a wave or a \emph{BLF}. For each $p \in V$, there exists a unique $p$-event in $\left[ C_{1},C_{2}\right] $
%which is minimal among the p-events in $\left[ C_{1},C_{2}\right] $, according to the order $\preceq $. We call it the initial event of $p$ in $\left[ C_{1},C_{2}\right] $.

%\end{lemma}
%\begin{proof}
%Let $(q,t)$ a decide event, its past in $\left[ C_{1},C_{2}\right] $ covers $V$, so there exists a $p$-event. 
%The set of $p$-events is a finite chaine, according to $\preceq $. The lemma follows.
%\end{proof}

\subsection{Infima and $r-$operators}

 Tel, in his work about wave algorithms \cite{Tel94}, introduces the infimum
operators. An infimum $\oplus $ over a set $\Bbb{S}$, is an associative,
commutative and idempotent (i.e. $x\oplus x=x$) binary operator. If $P=\left\{ a_{1},a_{2},...,a_{r}\right\} $ is a finite part of $\Bbb(S)$ then, from the associativity,
$\oplus P$ makes sens as $a_{1}\oplus a_{2}\oplus ...\oplus a_{r}$. And if $a\in S$, then $a \oplus P$ makes sens as $a\oplus a_{1}\oplus a_{2}\oplus ...\oplus a_{r}$.
Such an
operator defines a partial order relation $\leq _{\oplus }$ over $\Bbb{S}$,
by $x$ $\leq _{\oplus }y$ if and only if $x\oplus y=x$.\ We suppose that $\Bbb{S}$
has a greater element $e_{\oplus }$, such that $x$ $\leq _{\oplus }e_{\oplus }$
for every $x\in \Bbb{S } .$ Hence $\left( 
\Bbb{S } ,\oplus \right) $ is an Abelian
idempotent semi-group with $e_{\oplus }$ as identity element for $\oplus $. 
Ducourthial introduces in \cite{Duc98} the notion of $r$-operator which
generalizes the infimum operators.%  and still allows convergence to terminal
%configuration of wave-like algorithms.

\begin{definition}
The binary operator $\triangleleft $ on $\Bbb{S } $ is a $r-$operator if there exits a $\left( \Bbb{S } ,\oplus \right) $-endomorphism $r$, called $r$\emph{%
-function}, such that: 
$\forall x,y\in \Bbb{S} ,\ x\triangleleft
y=x\oplus r\left( y\right) $.
Let $\triangleleft $ be a $r-$operator on $\Bbb{S} $, and let $r$ be its associated $r-$\emph{function} , $%
\triangleleft $ is \emph{idempotent} if and only if: 
$\forall x\in \Bbb{S} ,\ x\leq _{\oplus }r(x)$.
%and $\triangleleft $ is strictly \emph{idempotent} if and only if: 
%$\forall x\in \Bbb{S } ,\ x<_{\oplus }r(x)$
A mapping $\triangleleft $ from $\left( \Bbb{S } \right) ^{n}$ to $\Bbb{S } $ is an $n-$ary 
$r-$operator if there exists $n-1$ $\left( \Bbb{S },\oplus \right) $-endomorphisms $r_{1},r_{2},...,r_{n-1}$
such that for all $\left( x_{0},x_{1},...,x_{n-1}\right) \in \left( \Bbb{S }\right) ^{n}:$
 $\triangleleft \left( x_{0},x_{1},...,x_{n-1}\right) =x_{0}\oplus r_{1}\left(
x_{1}\right) \oplus ...\oplus r_{n-1}\left( x_{n-1}\right) $

\end{definition}

\begin{remark}
\label{rem:r_compat}
$r$ is an endomorphism, which means that for all $x,y$ in $\Bbb{S }$, $r(x\oplus  y)=r(x)\oplus r( y)$. From the definition of $\leq _{\oplus }$,
we deduce that $r$ is compatible with $\leq _{\oplus }$, formally: $\forall x,y\in \Bbb{S }  ,x\leq _{\oplus }y\Rightarrow$ $ r\left( x\right)
\leq _{\oplus }r\left( y\right) $

\end{remark}

%\begin{remark}
%For every $x\in \Bbb{S\cup }\left\{ e_{\oplus }\right\} $, $r\left( x\right)
%=e_{\oplus }\oplus r(x)=e_{\oplus }\triangleleft x$, so $r$ is uniquely
%defined by $\triangleleft .$
%\end{remark}

%\begin{proposition}
%If $\triangleleft $ is a binary operator on $\Bbb{S }\cup\left\{ e_{\oplus
%}\right\} $, and $r$ is the $r-$operator associated, then:

 %$r-$associativity: $\forall x,y,z\in \Bbb{S }\cup\left\{ e_{\oplus
%}\right\} ,\ \left( x\triangleleft y\right) \triangleleft r\left( z\right)
%=x\triangleleft \left( y\triangleleft z\right) $

 %$r-$commutativity: $\forall x,y\in \Bbb{S}\cup \left\{ e_{\oplus
%}\right\} ,\ r\left( x\right) \triangleleft y=r\left( y\right) \triangleleft
%x$

 %$r-$idempotency: $\forall x\in \Bbb{S}\cup \left\{ e_{\oplus
%}\right\} ,\ r\left( x\right) \triangleleft x=r\left( x\right) $

%\end{proposition}

%Further forward, we will suppose that $\triangleleft $ is idempotent.

\subsection{Infimum and   $r$-operator parametrized computation problem}

 Let $\left[ C_{1},C_{2}\right] $ be a wave . We denote by $\mathcal{N}_{p}^{t}$ the set of processes  such that there exists a time $t_q$ such that $\left( q,t_{q}\right) \leadsto \left( p,t\right)$. Note that $p$ may be in $\mathcal{N}_{p}^{t}$. Because of the lack of place, the proof of Theorem~\ref{th:roperator} is in the annexe .

\paragraph{Infimum computation
}
Give each process $p$, an extra variable $p.res:\Bbb{S}$ . Each register $p.res$ is initialised during the initial event of $p$ by the value $p.v_0$. 
let $(p,t) $ be any   event in $\left[ C_{1},C_{2}\right] $. Whenever  $(p,t) $ holds,  
$p.res$ is set to the value $p.v_0\bigoplus \left\{ q^{t_{q}}.res,q \in \mathcal{N}_{p}^{t}\right\} $. Tel shows the following theorem:

\begin{theorem}~\cite{Tel94}
\label{th:infimum}
A wave can be used to compute an infimum.
\end{theorem}

%\begin{proof}
%%
%We claim that at the end of each event $(p,t)$ the register $p.res$ is equal to $\bigoplus \left\{ q.v_0,q\in Cover((p,t))\right\} $. 
%Let $\mathcal{A}$ be the set of events $(p,t)$ such that  $p.res \neq \bigoplus \left\{ q.v_0,q\in Cover((p,t))\right\} $.  If $\mathcal{A}$ is empty, the proof is finished. Note that the %minimal events in $\left[ C_{1},C_{2}\right] $ are not in $\mathcal{A}$. Suppose that
%$\mathcal{A}$ is not empty.  Let $(p,t)$ a minimal event of $\mathcal{A}$  according to the relation $ \preceq $. At time $t$, $p.res:=p.v_0\bigoplus \left\{ q.res,q\in %\mathcal{N}_{p}^{t}\right\} $. If $q\in \mathcal{N}_{p}^{t}$, we denote $t_{q}$ the time such that $\left(q,t_{q}\right) \leadsto \left( p,t\right) $, 
%then $Cover(p,t)=\left\{ p\right\} \bigcup\limits_{q\in \mathcal{N}_{p}^{t},}Cover\left(
%q,t_{q}\right) $. Because of the minimality of the event $(p,t)$, we deduce that for each $q \in \mathcal{N}_{p}^{t}$,   the equality $q.res = \bigoplus \left\{ s.v_0.,s\in %Cover((q,t))\right\} $ holds at time $t_q$. It results that $p.res = \bigoplus \left\{ q.v_0.,q\in Cover((p,t))\right\} $ and $\mathcal{A}$ is empty.
%We know that for any decide event $(p,t)$ of the sequence $\left[ C_{1},C_{2}\right] $,  $Cover(p,t)=V$. Thus, following the result above, 
% $p.v=\bigoplus \left\{ q.v_0.,q\in V \right\}$ at time $t$.
%
%\end{proof}

\paragraph{Idempotent $r$-operator parametrized  computation problem}

Let $\left[ C_{1},C_{2}\right] $ be a strong wave . 
We associate to each oriented link $\left( p_{i},p_{j}\right) $ of $G=\left(
V,E\right) $ a idempotent $r-$\emph{function}: $r_{p_{i},p_{j}}$. By extention, for the sequence $(p_i,p_i)$ we associate the identity: $r_{ii}=id$.
Like above, give each process $p$, an extra constant $p.v_0:\Bbb{S}$ and a register $p.res$.
Each register $p.res$ is initialised during  the initial event of $p$ by the value $p.v_0$. 
let $(p,t) $ be any   event in $\left[ C_{1},C_{2}\right] $. Whenever  $(p,t) $ holds,  
$p.res$ is set to the value 
$p.v_0 \oplus  \left\{ r_{q,p}\left( q.res\right), q\in  \mathcal{N}_{p}^{t}     \right\} $. Each node $p$ can be seen as a $(d+1)$-ary $r$-operator if $d$ is the degree of the node.

\begin{definition}

For any walk $\mu=p_{0}p_{1}....p_{n}$, we define $eval\left( \mu \right) =r_{\mu }\left(
p_0.v_{0}\right) $ , with \\ $r_{\mu }=r_{p_{n-1},p_{n}}or_{p_{n-2},p_{n-1}}o...or_{p_{0},p_{1}}$,  where $o$ is
the composition of functions.
For any  $ p \in V$, the sets $\Lambda _{p}^{\prime } $ and $\Lambda _{p}$ are defined by: 
$\Lambda _{p}^{\prime }  =\left\{ eval\left( \mu \right)
,\mu \in \Sigma _{p}^{\prime }\right\}  \text{ and }
\Lambda _{p} = \left\{ eval\left( \mu \right) ,\mu \in
\Sigma _{p}\right\} $, where $\Sigma _{p}^{\prime }$ is the set of the\emph{\ walks} ending to $p
$, and 
$\Sigma _{q}$ is the set of the \emph{simple walks} ending to $p
$.

\end{definition}

From the definitions and the idempotence of the $r$-operators, the following lemma holds:
\begin{lemma}
\label{lem:reduc_walk}
 Assume that $m$ and $m'$ are two walks with $p=head(m)$. We suppose that $m\stackrel{*}{\rightarrow }m^{\prime }$. 
Then $r_{m }(p.v_0)\geq_{\oplus } r_{m{\prime } }(p.v_0)$, and if $m$ is elementary then  $r_{m }(p.v_0)=r_{m{\prime } }(p.v_0)$ 
\end{lemma}

\begin{definition}[Legitimate output]
We define the legitimate output of a process $p$ as the quantity: 
$\oplus \Lambda _{p}$

\end{definition}

\begin{theorem}
\label{th:roperator}
A \emph{strong wave} can be used to solve the idempotent $r$-operator parametized problem.
\end{theorem}

\section{Unison as a self-stabilizing wave stream algorithm, applications }

\label{sec:applications}

\subsection{Analysis of the Unison Behavior starting in $WU_0$}

\begin{lemma}
Let $k\geq \bot_0 $.\ If   $\left( p,t\right) $ is an event in  the interval  
$\left[ C_{k},\rightarrow \right[ $, then: $V(p,\widetilde{p^{t}.r}-k)\subset
Cover(p,t)$ and $\Sigma _{p}^{\widetilde{p^{t}.r}-k}\subset WalkCover(p,t)$.

Where $\Sigma _{p}^{\rho }$ is the set of simple walks of length less than
or equal to $\rho $, ending to $p$. 
\end{lemma}

\begin{proof}
\small
The lemma is true for the initial events of $\left[ C_{k},\rightarrow
\right[ $. Let $\mathcal{A}$ be the set of events $(p,t)$ in  $\left[
C_{k},\rightarrow \right[ $ such that the sentence: 
$$
V(p,\widetilde{p^{t}.r}-k)\subset
Cover(p,t)\  \wedge \ \Sigma _{p}^{\widetilde{p^{t}.r}-k}\subset
WalkCover(p,t)
$$ 
does not hold.  We assume that $\mathcal{A}$ is not empty, let $(q,\tau )$ a minimal event in $\mathcal{A}$ according to $\preceq $. 
Let $\delta =\widetilde{q^{\tau }.r}-k$, and let $p_{1}\in V(q,\delta )$. If $p_{1}=q$ then $p_{1}\in Cover(q,\tau )$, else there exists $q_{1}\in 
\mathcal{N}_{q}$ such that $p_{1}\in V(q_{1},\delta -1).$ $(q,\tau )$is not
a initial event, so $q_{1}\in \mathcal{N}_{q}^{\tau }$ and there exists $%
\tau _{q_{1}}$ such that $\left( q_{1},\tau _{q_{1}}\right) \leadsto $ $%
\left( q,\tau \right) $, and by the minimality of $(q,\tau )$ the inclusion $%
V(q_{1},\delta -1)\subset Cover(q_{1},\tau _{q_{1}})$ holds and thus $%
V(q_{1},\delta -1)\subset Cover(q,\tau )$ and $p_{1}\in Cover(q,\tau )$.
Following the same way, let $m$ be a walk in $\Sigma _{p}^{\delta }$ , if $%
m=q$ then $m\in $ $WalkCover(q,\tau )$, else if $m=p_{1}p_{2}....p_{r}q$
then  $p_{r}\in \mathcal{N}_{q}^{\tau }$ because$(q,\tau )$ is not a initial
event, so  there exists $\tau _{p_{r}}$ such that $\left( p_{r},\tau
_{p_{r}}\right) \leadsto $ $\left( q,\tau \right) $, and by the minimality
of $(q,\tau )$ the inclusion $\Sigma _{p_{r}}^{\delta -1}\subset
WalkCover(p_{r},\tau _{p_{r}})$ holds, and thus $p_{1}p_{2}....p_{r}q\in
WalkCover(q,\tau )$ . So $(q,\tau )$ is not in $\mathcal{A}$. Thus  $%
\mathcal{A=}\emptyset $, and the lemma is proved.
\normalsize
\end{proof}

As corollary, we deduce the important following theorem:

\begin{theorem}
\label{th:unison_behavior}
Let $k\geq \bot_0 $ and $\delta $ be a positive integer, then $\left[
C_{k},C_{k+\delta }\right] $, with $C_{k+\delta }$ as the set of decide events, is a $\delta $\emph{-wavelet}, and a \emph{wave}
if $\delta \geq D$. If $\delta $ is greater than or equal to the length of
a longest simple walk in $G$, then $\left[ C_{k},C_{k+\delta }\right] $
is a strong wave.
\end{theorem}

\subsection{Self-stabilizing computation of an infimum at distance  $\rho$}

\paragraph{If $\rho \geq D$}
For each process $p$, the registers $p.v_{0}$ and $p.res$ are  intializedby
the same value. We need one step for the initilisation, and $D$ steps for
the wave of calculation. So we take $\delta \geq D+1$. For any integer $U$, $\left[ C_{U\delta },C_{U\delta +\delta -1}\right] $ is a wave. So we define
the critical sections as follows:
\small
$$
<<CS2>>\equiv \text{ initialization of }p.v_{0}\text{ and }p.res  \\
$$
$$
<<CS1>>\equiv p.res:=p.v_{0}\bigoplus \left\{ q.res,q\in \mathcal{N}_{p}\right\} 
$$
\normalsize
From Theorem~\ref{th:infimum},  at the cut $C_{U\delta +\delta -1}$ the register $p.res$ contains the right value $\bigoplus \left\{ q.v_{0},q\in V\right\} $.

\paragraph{If $\rho < D$}

We take $\delta =\rho +1$.\
We suppose that the register $q.v_{0}$ is initialised during the critical
section $<<CS2>>$ at the beginning of the phase, precisely when the register 
$p.r$ takes the value $U\delta $.
To reach the objective, we define for each process $p$ two added registers 
$p.v_{1}$ and $p.v_{2}$. These two registers are initialized  at the date $C_{U\delta }$ during the critical section $<<CS2>>$, by the value $p.v_{0}$.
For $\alpha \in \left\{ 1,2,...,\rho \right\} $, at the date $C_{U\delta
+\alpha }$, the action $<<CS1>>$ is the following:
\small
$$
p.v_{1}:=p.v_{2};p.v_{2}:=p.v_{0}\bigoplus \left\{ q.v_{\varphi \left(
q\right) },q\in \mathcal{N}_{p}\right\} 
$$
\normalsize
with,  if $q.r=p.r$ then  $\varphi \left( q\right) =2$, and if $q.r=p.r+1$  
then $\varphi \left( q\right) =1$.

\begin{theorem}
\label{th:rho_min}
 At the cut  $C_{U\delta +\delta -1}$ the register $p.res$ contains the right value: $\bigoplus \left\{ q.v_{0},q\in V (p,\rho)\right\} $.
\end{theorem}
The proof is in the annexe.

\section{Concluding remarks}
\label{sec:conclusion}

We showed how the  stucture of the communications between processes of Unison can be viewed as a wave stream.
Thanks to this structure, we have been able to build a self-stabilizing wave stream algorithm in asynchronous anonymous
networks scheduled by an unfair daemon.  
Precisely, we showed that the behavior of Unison can be viewed as a self-stabilizing
wave, $k$-wavelet or bidirected link flood streams. 
From these remarks, in any asynchronous anonymous network scheduled by an unfair daemon, we deduced
  self-stabilizing solutions to the barrier synchronization problem,
 the infimum calculation problem,
and   the idempotent
$r$-operator parametrized calculation problem.
Now, an important question would be to reduce the self-stabilizing time complexity  of  unison from $O(n)$ to $O(D)$ in a general graph.

\begin{footnotesize}
\bibliographystyle{alpha}
\bibliography{unison}
\end{footnotesize}

\section{Annexe}

\subsection{Proof of  Theorem~\ref{th:roperator}}
\begin{proposition}
For any  $ p \in V$, $\oplus \Lambda _{p}^{\prime
}$ exists, and the equality $\oplus \Lambda _{p}^{\prime
} =\oplus \Lambda _{p} $
holds. 
\end{proposition}

\begin{proof}

$\Lambda _{p}$ is finite, so $\oplus \Lambda _{p}$ exists, furthermore $\Lambda _{p}  \subset  \Lambda _{p}^{\prime} $. 
If $m =p_{0}p_{1}....p_{n}$ be a not elementary walk , there exists a simple walk $m^{\prime}$  such that $m\stackrel{*}{\rightarrow }m^{\prime }$. 
From the lemma~\ref{lem:reduc_walk}, $r_{m }(p.v_0)\geq_{\oplus } r_{m{\prime } }(p.v_0)$ and $r_{m{\prime } }(p.v_0) \geq_{\oplus } \oplus \Lambda _{p}$ hold. The proposition follows.

\end{proof}

\begin{lemma}
\label{lem:etat}
Let $(p,t)$ be an event in $\left[ C_{1},C_{2}\right] $, then at time $t$,

$$p^t.res=\bigoplus \left\{ eval\left( \mu \right) ,\mu \in WalkCover(p,t)\right\} $$

\end{lemma}

\begin{proof}

Let $\mathcal{A}$ be the set of events $(p,t)$ in $\left[ C_{1},C_{2}\right] $ such that the equality is not true.  Note that the minimal events in $\left[ C_{1},C_{2}\right] $ are not in $\mathcal{A}$.
If $\mathcal{A}$ is empty, the proof is finished. Suppose that
$\mathcal{A}$ is not empty.  Let $(p,t)$ a minimal event of $\mathcal{A}$  according to the relation $ \preceq $:
%For $q\in \mathcal{N}_{p}^{t}$, we denote $t_{q}$ the time such that $\left(q,t_{q}\right) \leadsto \left( p,t\right) $.
 $$
p^t.res:=p.v_0\bigoplus \left\{ r_{q,p}(q^{t_q}.res),q\in \mathcal{N}_{p}^{t}\right\} 
$$
 But, by definition: 
$
WalkCover(p,t)=\bigcup\limits_{q\in \mathcal{N}_{p}^{t}}\left\{ \mu p,\mu
\in WalkCover\left( q,t_{q}\right) \right\} \cup \left\{ p\right\}
$. 

From the minimality of $(p,t)$ in $\mathcal{A}$, the events $(q,t_q)$ are not in $\mathcal{A}$, so:
$$
p^{t}.res=p.v_{0}\bigoplus\limits_{q\in \mathcal{N}_{p}^{t}}r_{qp}\left(\bigoplus \left\{ eval\left( \mu \right) ,\mu \in
WalkCover\left( q,t_{q}\right) \right\} \right) 
$$
But $r_{pq}$ is compatible with $\leq $ $_{\bigoplus }$ (remark~\ref{rem:r_compat}), thus:
$$
p^{t}.res=p.v_{0}\bigoplus\limits_{q\in \mathcal{N}_{p}^{t}}
\left\{ r_{qp}(eval(\mu)) ,\mu \in
WalkCover\left( q,t_{q}\right) \right\} 
$$
 $(p,t)$ is not an initial event, so $p\in \mathcal{N}_{p}^{t}$ and:
$$
p.v_{0}\geq _{\oplus }\bigoplus \left\{ r_{pp}or_{\mu }\left( head(\mu
).v_{0}\right) ,\mu \in WalkCover\left( p,t_{p}\right) \right\} 
$$
We deduce,
from associativity of $\oplus $ and from $r_{qp}or_{\mu }=r_{\mu p}$ that :

$$
p^{t}.res=\bigoplus\limits_{q\in \mathcal{N}_{p}^{t}}\left\{ eval(\mu p)  ,\mu \in WalkCover\left( q,t_{q}\right) \right\} 
$$
but $WalkCover(p,t)=\bigcup\limits_{q\in \mathcal{N}_{p}^{t}}\left\{ \mu
p,\mu \in WalkCover\left( q,t_{q}\right) \right\} $, so: 
$$
p^{t}.res=\left\{ eval(\mu)  ,\mu \in
WalkCover\left( p,t\right) \right\} 
$$
We deduce that $(p,t)$ is not  in $\mathcal{A}$, which is a contradiction. We deduce that $\mathcal{A}=\emptyset$ and the lemma.

\end{proof}

%where $\Lambda _{p}=\bigcup\limits_{q\in V}\Lambda _{p}\left( q\right) .$

%\begin{remark}
%For all $p\in V$, $\Lambda _{p}\left( p\right) =\left\{ p.v\right\} $
%\end{remark}

\begin{theorem}[\ref{th:roperator}]

A \emph{strong wave} can be used to solve the idempotent $r$-operator parametized problem.
\end{theorem}

\begin{proof}
\small
If $\left( p,t\right) $ is a decide event then, from the Lemma~\ref{lem:etat}, we etablish that
$p^{t}.res=\left\{ eval\left(\mu \right) ,\mu \in WalkCover\left( p,t\right) \right\} $ holds, 
 and that $WalkCover\left( p,t\right) $ satisfies the Definition~\ref{def:wave}. 
Recall that $\Lambda _{p}=\left\{ eval\left( \mu \right) ,\mu \in \Sigma _{p}\right\} $.
For any $m\in WalkCover\left( p,t\right) $, there exists $m_{0}\in \Sigma_{p}$ such that 
$m\stackrel{*}{\rightarrow }m_{0}$ and from the Lemma~\ref{lem:reduc_walk} the
inequality $eval\left( m\right) \geq eval\left( m_{0}\right) $ holds. We
deduce that $p^{t}.res\geq \oplus \Lambda _{p}.$
Conversely, if $m_{0}\in \Lambda _{p}$, there exists $m\in WalkCover\left(
p,t\right) $ such that $m\stackrel{*}{\rightarrow }m_{0}$, but from the Lemma~\ref{lem:elt_walk_red}, 
 there exists also a walk $m_{1}\in WalkCover\left( p,t\right) $ such that 
$m\stackrel{*}{\rightarrow }m_{1}$ and $m_{1}\stackrel{*}{\rightarrow }m_{0}$,
and from Lemma~\ref{lem:elt_walk_red} $eval\left( m_{1}\right) =eval\left( m_{0}\right) $. We
deduce that $p^{t}.res\leq \oplus \Lambda _{p}$.
From these two inequalities, we deduce $p^{t}.res=\oplus \Lambda _{p}$ and the
theorem is proved.
\normalsize
\end{proof}

\subsection{Proof of  Theorem~\ref{th:rho_min}}

\begin{proposition}
For $p \in V$ and $\alpha \in \left\{ 1,...,\rho \right\} $, at
the date $C_{U\delta +\alpha }$, hold the equalities:

$$
p.v_{1}=\bigoplus \left\{ q.v_{0},q\in V\left( p,\alpha -1\right) \right\} 
\text{ and }p.v_{2}=\bigoplus \left\{ q.v_{0},q\in V\left( p,\alpha \right)
\right\} 
$$

\end{proposition}

\begin{proof}

At the date   $C_{U\delta },$ any process $p$ satisfies $p.v_{1}=p.v_{0}$
and $p.v_{2}=p.v_{0}$, it is the initializing step. Let $\mathcal{A}$ the set
of events in $\left[ C_{U\delta +1},C_{U\delta +\delta -1}\right] $, for
which the proposition is not true. We assume that $\mathcal{A}$ is not
empty. Let $(p,t)$ a minimal event in $\mathcal{A}$. let $\alpha \in \left\{
1,2,...,\rho \right\} $ such that $(p,t)$ $\in C_{U\delta +\alpha }$.
There exists $t_{0}$ such that $\left( p,t_{0}\right) \rightsquigarrow $ $%
(p,t)$. We have $p^{t}.v_{1}=p^{t_{0}}.v_{2}=$ and $p^{t_{0}}.v_{2}=\bigoplus
\left\{ q.v_{0},q\in V\left( p,\alpha -1\right) \right\} $. This equality is
true even if $\alpha =1$.
Now, $p^{t}.v_{2}=p.v_{0}\bigoplus \left\{ q^{t_{q}}.v_{\varphi \left(
q\right) },q\in \mathcal{N}_{p}\right\} $. From the minimality of the event $%
\left( p,t\right) $, the events $\left( q,t_{q}\right) $, where $t_{q}<t$, 
are not in $\mathcal{A}$ and are in $\left[ C_{U\delta },C_{U\delta +\delta
-1}\right] $. 
So, $p.v_{0}\bigoplus \left\{ q^{t_{q}}.v_{\varphi \left(
q\right) },q\in \mathcal{N}_{p}\right\} =\bigoplus \left\{ q.v_{0},q\in
V\left( p,\alpha \right) \right\} $. We obtain a contradiction. We
deduce that $\mathcal{A}$ is empty, and the proposition follows

\end{proof}

As corollary, we obtain the theorem:

\begin{theorem}[\ref{th:rho_min}]
 At the cut $C_{U\delta +\delta -1}$ the register $p.res$ contains the right value $\bigoplus \left\{ q.v_{0},q\in V (p,\rho)\right\} $.
\end{theorem}

\end{document}